\documentclass[amsmath,11pt]{article}

\usepackage{amssymb,amssymb,epsfig,bm}
\usepackage{caption}
\usepackage{subcaption}

\date{\empty}

\begin{document}

\title{\bf Peculiar velocities in the early universe}
\author{Myrto Maglara${}^1$ and Christos G. Tsagas${}^{1,2}$\\ {\small ${}^1$Section of Astrophysics, Astronomy and Mechanics, Department of Physics}\\ {\small Aristotle
University of Thessaloniki, Thessaloniki 54124, Greece}\\ {\small ${}^2$Clare Hall, University of Cambridge, Herschel Road, Cambridge CB3 9AL, UK}}

\maketitle

\begin{abstract}
Large-scale peculiar motions are commonplace in our universe. Nevertheless, their origin, evolution and implications are still largely unknown. It is generally assumed that bulk motions are a relatively recent addition to the universal kinematics, triggered by the increasing inhomogeneity and anisotropy of the post-recombination epoch. In this work, we focus on the linear evolution of peculiar velocities prior to recombination, namely in the late radiation era and also during a phase of de Sitter inflation. We begin by showing/confirming that bulk motions are triggered and sustained by the non-gravitational forces developed during structure formation. Since density and therefore peculiar-velocity perturbations cannot grow in the baryonic sector before recombination, we consider drift motions in non-baryonic species, which can start growing in the late radiation era. Using relativistic linear cosmological perturbation theory, we find that peculiar motions in the low-energy dark component exhibit power-law growth, which increases further after equilibrium. Turning to the very early universe, we consider the evolution of linear peculiar velocities during de Sitter inflation. We find that typical slow-roll scenarios do not source peculiar motions. Moreover, even if the latter were to be present at the onset of the de Sitter phase, the subsequent exponential expansion should quickly wash away any traces of peculiar-velocity perturbations.
\end{abstract}

\section{Introduction}\label{sI}
Observations have repeatedly verified the presence of large-scale peculiar motions, commonly referred to as bulk flows, with typical sizes of few hundred Mpc and velocities of few hundred km/sec~\cite{WFH}-\cite{Saetal}. The various surveys seem to (roughly) agree on the direction of these motions, but they generally do not agree on their size and speed. Perhaps the most extreme example are the so-called ``dark flows'', with sizes and velocities well in excess of those typically anticipated~\cite{KA-BKE}. Although such dark flows appear to be at odds with the theoretical predictions of the current cosmological paradigm, as well as with the Planck data~\cite{Aetal}, the truth of the matter is that the origin, the evolution and the implications of the observed peculiar motions remain largely unknown. The aim of this work is to look at the bulk-flow question from the theoretical point of view. In so doing, we assume a ``tilted'' perturbed Friedmann-Robertson-Walker (FRW) universe and employ relativistic linear cosmological perturbation theory (e.g.~see~\cite{TCM,EMM} and references therein). Tilted cosmologies are fairly straightforward generalisations of perturbed cosmological spacetimes, allowing for two (at least) families of relatively moving observers. In this respect, the titled models provide a more accurate representation of the actual universe and are therefore better suited for the study of the observed large-scale peculiar motions and of their implications.

Most of the available theoretical studies of cosmological peculiar motions, which are rather few and sparse to begin with, are Newtonian in nature (e.g.~see~\cite{Pe}-\cite{Pa}). The reason can be probably traced in the widespread perception that any relativistic corrections that may emerge should be too weak to make a noticeable difference. For instance, the velocities of the reported peculiar motions are well below the relativistic limit. Also, on the scales of interest, the  curvature effects are too weak to have a measurable effect. Nevertheless, Newtonian gravity and general relativity are fundamentally different in their treatment of both the gravitational field itself and of its sources. More specifically, only the density of the matter contributes to the Newtonian gravitational potential, via Poisson's equation. This is not the case in Einstein's theory, where the pressure, the energy-flux and the viscosity of the matter also contribute to the gravitational field, through the energy-momentum tensor. When dealing with bulk peculiar flows, there is always a nonzero energy-flux input to the local gravitational field at the linear perturbative level, namely even when the involved velocities are non-relativistic. In practice, this means that, even when the cosmic medium is perfect, it must be treated as (effectively) imperfect, if the linear relativistic effects of peculiar motions are be to accounted for (see relation (\ref{lrels1}c) in \S~\ref{ssARM} below). Although this theoretical argument is not entirely new, it was only recently that its implications for the evolution and for the role of the observed large-scale peculiar motions were investigated (e.g.~see~\cite{TT,T1}).

Without accounting for the aforementioned gravitational input of the bulk-flow flux, linear peculiar velocities ($v$), on a homogeneous and isotropic Friedmann background, were found to grow at the rather moderate rate of $v\propto t^{1/3}$ after equipartition~\cite{Pe,M}. In fact, there have also been linear studies involving peculiar velocities that decay as $v\propto t^{-2/3}$~\cite{CMV-R} after equipartition. This seems at odds with the observations, all of which report the presence of extensive bulk peculiar motions in the universe. In all of the above studies, the flux contribution to the local gravitational field was (inadvertently) bypassed, or switched off. In contrast, when included in the linear analysis, the energy flux of the bulk motion lead to a considerably stronger growth rate, with $v\propto t^{4/3}$~\cite{TT,FT}. Technically speaking, this happens because the flux input to the energy-momentum tensor feeds into the relativistic conservation laws and eventually modifies the formulae monitoring the evolution of linear peculiar velocities. Intuitively, one could argue that bulk peculiar flows (in a sense) gravitate and the increase of their velocity growth is the result of the enhanced local gravitational field. The whole flux-effect is purely relativistic, with no known Newtonian analogue.

Large-scale bulk motions are believed to be a relatively recent addition to the kinematics of our universe, triggered by the ongoing process of structure formation. Prior to recombination, during the radiation era for example, density perturbations in the photon-baryon system cannot grow on scales smaller than the Jeans length. At the time, the latter essentially coincides with the Hubble horizon. Also, between equilibrium and decoupling, baryonic density perturbations with wavelengths smaller than the Silk scale are erased by photon diffusion. As a result, peculiar velocities in the baryonic sector are not expected to develop before recombination, at least on subhorizon scales. Cold Dark Matter (CDM) particles, on the other hand, interact very weakly (if at all) with the rest of the species and their typical free-streaming scales are too small to be of cosmological interest. Nevertheless, perturbations in the CDM density are believed to remain stagnant during most of the radiation era due to the ``Meszaros effect''. Hence, inhomogeneities in the dark-matter begin to grow close to the equilibrium time, so it is only then that they can induce peculiar velocities in the CDM sector. Following~\cite{TT}, linear peculiar velocities in the baryonic and the (cold) dark matter components of the cosmic medium grow as $t\propto t^{4/3}$, after recombination and equipartition respectively. Here, we extend that study to analyse linear peculiar motions in the dark-matter sector during the radiation era, assuming that the species in question become non-relativistic before matter-radiation equality. We find that prior to equilibrium peculiar velocities in the CDM also grow, though at a rate slightly slower ($v\propto t^{(3+\sqrt{41})/8}$) than in the subsequent dust era. Nevertheless, after recombination, any pre-existing peculiar-velocity perturbations in the CDM sector should amplify peculiar motions in the baryons, in the same way that ``gravitational wells''  in the dark-matter distribution accelerate the growth of baryonic density perturbations after decoupling.

There have been speculations in the literature that the peculiar-velocity fields observed today might have primordial origin~\cite{T}. If true, such a possibility would inevitably have pivotal implications for our understanding of the early universe, as well as of the way it has evolved to the present. With this in mind, we consider the linear evolution of peculiar velocities during inflation, by allowing for two 4-velocity fields moving relative to each other. In so doing, we assume exponential de Sitter inflation, driven by a minimally coupled scalar field. The key difference with inflation is that the effective equation of state of a slow-rolling scalar field (i.e.~the fact that $p=-\rho$) ensures that, if the cosmic medium is perfect, the linear energy-flux vector vanishes in all frames (see relation (\ref{lrels1}c) in \S~\ref{ssARM} below), despite their relative motion. In the absence of the flux input, it becomes impossible to generate peculiar-velocity perturbations at the linear level. Moreover, any 4-velocity tilt that might have been present at the onset of the de Sitter phase, is completely washed away by the exponential expansion.  Although it is conceivable that more sophisticated inflationary scenarios could allow peculiar velocities to survive, our results do not seem to favour a primordial origin for them.

The exponential decay of peculiar-velocity perturbations during de Sitter inflation is not a surprise. What is worth noting is the entirely different fate of peculiar velocities after inflation (when their flux contributes to the local gravitational field) and during the de Sitter phase (when the gravitational input of the flux vanishes altogether). The difference emphasises the key role and the importance of the peculiar-flow flux for the evolution of cosmological peculiar motions.

In closing, we should note that one can perform the analysis in the tilted frame of the real observers, or in the reference coordinate system of their idealised (Hubble-flow) counterparts. Here, we will take both perspectives (see Fig.~\ref{fig:pmotion} in \S~\ref{ssARM} next) and reach the same results. This should not come as a surprise, of course, since physics is not frame-dependent.

\section{Cosmological peculiar-velocity fields}\label{sCP-VFs}
Real observers do not follow the smooth cosmic expansion, but we all have finite peculiar velocities with respect to it. As a result, the kinematics and the dynamics of the host universe, experienced by these relatively moving observers, generally differ.

\subsection{Aspects of relative motion}\label{ssARM}
\begin{figure}[!tbp]
  \begin{subfigure}[b]{0.475\textwidth}
    \includegraphics[width=\textwidth]{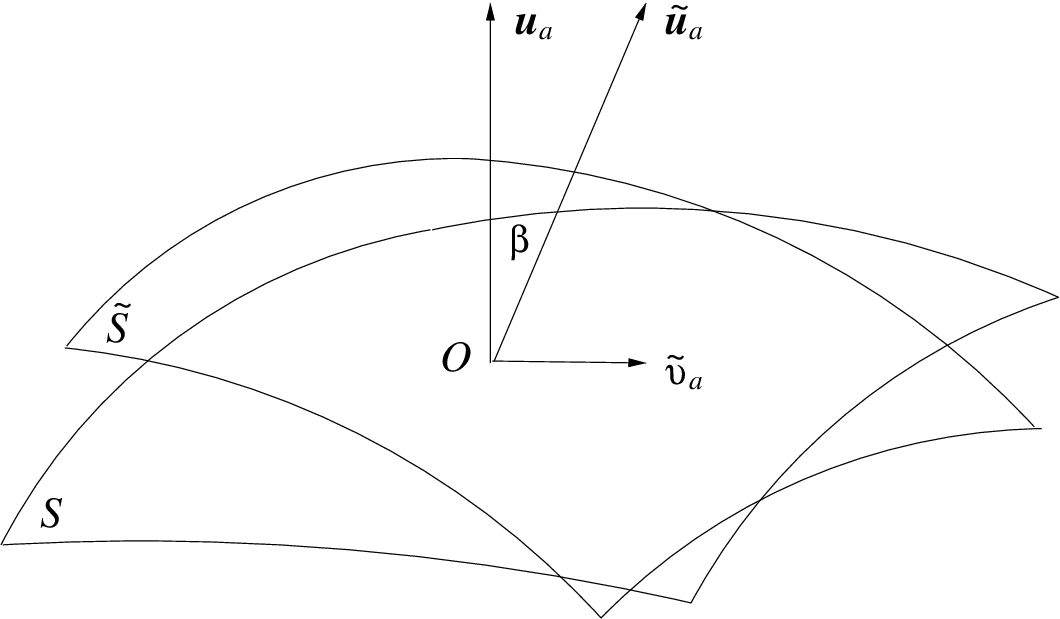}
    \caption{Tilted-frame perspective.}
    \label{fig:f1}
  \end{subfigure}
  \hfill
  \begin{subfigure}[b]{0.475\textwidth}
    \includegraphics[width=\textwidth]{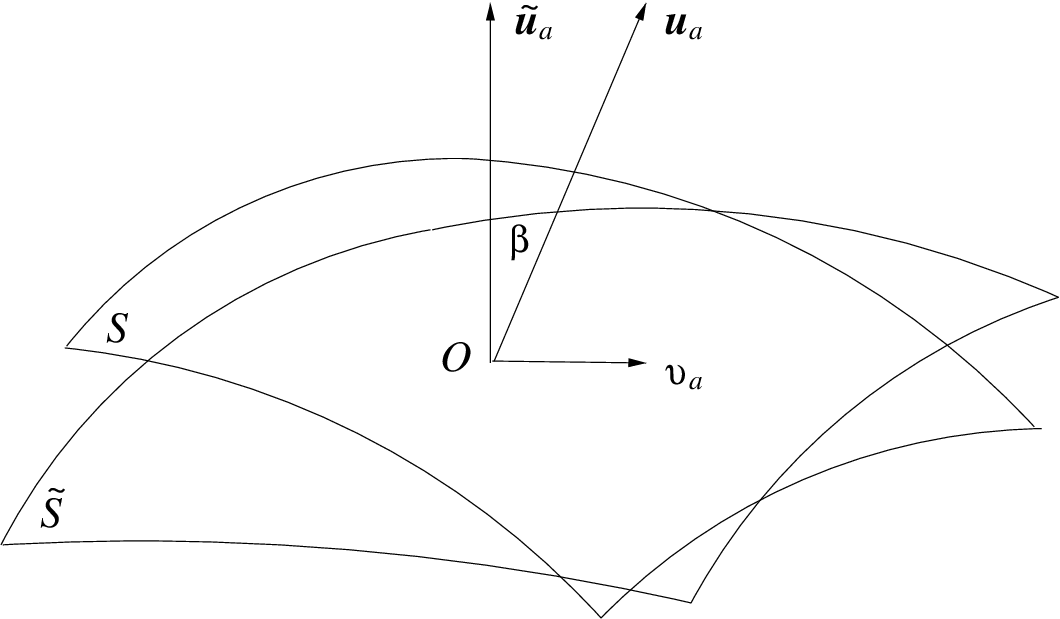}
    \caption{Hubble-frame perspective.}
    \label{fig:f2}
  \end{subfigure}
  \caption{Tilted spacetimes allow for two families of relatively moving observers, with 4-velocities $u_a$ and $\tilde{u}_a$, at every event ($O$). Assuming that the $u_a$-field defines the reference frame of the universe, $\tilde{u}_a$ is the 4-velocity of the real observers, ``drifting'' with peculiar velocity $\tilde{v}_a$ relative to the Hubble expansion (see Eq.~(\ref{4vels}a) and Fig.~\ref{fig:f1} above). Alternatively, one may turn to Fig.~\ref{fig:f2} and adopt the viewpoint of the idealised Hubble-flow observers. In that case, the latter are assumed to ``move'' relative to the real observers living in a typical galaxy with (effective) ``peculiar'' velocity $v_a$. It goes without saying that both approaches are physically equivalent and both lead to the same results (e.g.~see~\cite{TT,T1} and also \S~\ref{sCP-VFs}, \S~\ref{sPVPM} and \S~\ref{sPVDI} here). Note that in either case $\beta$ (with $\cosh\beta=-u_a\tilde{u}^a$) is the hyperbolic (tilt) angle between $u_a$ and $\tilde{u}_a$ (e.g.~see~\cite{TCM,EMM}). Also, $S$ and $\tilde{S}$ are the 3-D rest-spaces of the aforementioned two groups of observers.}  \label{fig:pmotion}
\end{figure}

In relativistic cosmology, the study of peculiar flows requires the use of the so-called tilted spacetimes, which allow for two (or more) groups of observers in relative motion with each other. Therefore, hereafter, we will consider a tilted almost-FRW universe and associate the aforementioned two families of observers with the 4-velocity fields $u_a$ and $\tilde{u}_a$ respectively. Then, when the peculiar motion is non-relativistic, we have
\begin{equation}
\tilde{u}_a= u_a+ \tilde{v}_a \hspace{15mm} {\rm and} \hspace{15mm} u_a= \tilde{u}_a+ v_a\,,  \label{4vels}
\end{equation}
with $\tilde{v}_a=-v_a$ at the linear perturbative level (e.g.~see~\cite{TCM,EMM} and also Fig.~\ref{fig:pmotion} here). Note that $\tilde{v}_a$ is the peculiar velocity of the tilted observers with respect to their non-tilted counterparts, whereas $v_a$ is the velocity of the latter group relative to the former (with $u_a\tilde{v}^a=0=\tilde{u}_av^a$ by construction).

Both of the frames defined above ``live'' in the perturbed almost-FRW spacetime. We assume that $u_a$-field defines the reference coordinate system the universe, typically identified with the idealised CMB-frame, where the dipole of the Cosmic Microwave Background (CMB) vanishes and with respect to which one can measure peculiar velocities. Then, $\tilde{u}_a$ is the 4-velocity of the real observers, living in galaxies like our Milky Way and moving with peculiar velocity $\tilde{v}_a$ relative to the Hubble flow (see Fig.~\ref{fig:f1} above). Alternatively, $v_a$ may be seen as the ``peculiar'' velocity of the Hubble frame with respect to the tilted observers (see Fig.~\ref{fig:f2}). Note that the CMB-frame is also commonly referred to as the Hubble-frame, so here we will use these two terms interchangeably. In the unlikely case that there is no reference frame in the universe, one can no longer talk about peculiar velocities. Then, the following analysis still holds, although now both 4-velocity fields correspond to real observers moving with respect to each other.

Relatively moving observers experience difference versions of ``reality'', even when the velocities involved are non-relativistic. For instance, at the linear level, the matter variables measured by the aforementioned observers in their own coordinate systems are related by~\cite{M}
\begin{equation}
\tilde{\rho}= \rho\,, \hspace{15mm} \tilde{p}= p\,, \hspace{15mm} \tilde{q}_a= q_a- (\bar{\rho}+\bar{p})\tilde{v}_a \hspace{10mm} {\rm and} \hspace{10mm} \tilde{\pi}_{ab}= \pi_{ab}\,,  \label{lrels1}
\end{equation}
Here, $\rho$, $p$, $q_a$ and $\pi_{ab}$ respectively represent the energy density, the (isotropic) pressure, the energy flux and the viscosity of the matter, as measured in the non-tilted (reference) frame, while $\tilde{\rho}$, $\tilde{p}$, $\tilde{q}_a$ and $\tilde{\pi}_{ab}$ are the same variables in the (tilted) coordinate system of the real observers. Al of the above are first order quantities measured in the perturbed universe. On the other hand, $\bar{\rho}$ and $\bar{p}$ (see (\ref{lrels1}c) are measured in the FRW background, where $\rho=\tilde{\rho}=\bar{\rho}$ and $p=\tilde{p}=\bar{p}$ by default. Following (\ref{lrels1}b) and (\ref{lrels1}d), if the pressure is zero in one frame, it vanishes in any other relatively moving coordinate system as well (to first approximation). This is not the case for the energy flux, however, since zero flux in one frame does not necessarily guarantee the same in the other (see Eq.~(\ref{lrels1}c) above). Moreover, recasting the latter expression as
\begin{equation}
(\bar{\rho}+\bar{p})\tilde{v}_a= q_a- \tilde{q}_a\,,  \label{tv1}
\end{equation}
it becomes evident that peculiar velocities require a nonzero energy-flux difference between the two frames. When the cosmic medium is perfect, we are free to set one of the two flux vectors on the right-hand side of (\ref{tv1}) to zero. Then, there is an effective energy flux in any other relatively moving coordinate system, simply because of the observers' peculiar motion. For instance, setting $q_a=0$ in the Hubble/CMB frame, implies that $\tilde{q}_a= -(\bar{\rho}+\bar{p})\tilde{v}_a\neq0$ in its tilted counterpart (unless $\bar{\rho}+\bar{p}=0$ -- see~\S~\ref{sPVDI} below). Alternatively, one may set $\tilde{q}_a=0$, in which case Eq.~(\ref{tv1}) ensures that $q_a=-(\bar{\rho}+\bar{p})v_a\neq0$ (recall that $v_a=-\tilde{v}_a$ to first approximation).

The role of the effective energy flux (either as $\tilde{q}_a$ or as $q_a$) in the evolution of peculiar velocities is pivotal, because it contributes to the (perturbed) energy-momentum tensor and therefore to the local gravitational field. In a sense, one could argue that bulk peculiar flows gravitate (see~\cite{TT,FT} for further discussion). The gravitational input of the peculiar flux is purely general relativistic in nature, it occurs at the linear perturbative level and it cannot be naturally reproduced in any Newtonian treatment. Most of the (few) relativistic treatments of cosmological peculiar motions also bypass the above mentioned flux contribution to the perturbed energy-momentum tensor. This happens because these studies either introduce an effective gravitational potential and thus become Newtonian in practice, or adopt the so-called energy (or Landau-Lifshitz) frame, where the flux vanishes by construction (e.g.~see~\cite{CMV-R} as well as~\cite{TCM,EMM}). In what follows, we will demonstrate that the growth, or decay, of cosmological peculiar velocities depends crucially on the presence, or the absence, of a nonzero peculiar flux (see \S~\ref{sPVPM} and \S~\ref{sPVDI} respectively).

Turning to the kinematic variables, as measured in the two coordinate systems defined above, we have the linear relations
\begin{equation}
\tilde{\Theta}= \Theta+ \tilde{\vartheta}\,,  \hspace{15mm} \tilde{\sigma}_{ab}= \sigma_{ab}+ \tilde{\varsigma}_{ab}\,, \hspace{15mm} \tilde{\omega}_{ab}= \omega_{ab}+ \tilde{\varpi}_{ab}  \label{lrels2a}
\end{equation}
and
\begin{equation}
\tilde{A}_a= A_a+ \tilde{v}_a^{\prime}+ H\tilde{v}_a\,,  \label{lrels2b}
\end{equation}
where the primes denote time differentiation in the tilted frame and $H$ is the Hubble parameter of the FRW background~\cite{M}. Here, $\Theta$, $\sigma_{ab}$, $\omega_{ab}$ and $A_a$ are respectively, the expansion scalar, the shear tensor, the vorticity tensor and the 4-acceleration vector measured in the (idealised) reference frame, with their tilted counterparts measured by the real observers. In addition, $\tilde{\vartheta}$, $\tilde{\varsigma}_{ab}$ and $\tilde{\varpi}_{ab}$ are the volume scalar, the shear and the vorticity tensors of the peculiar flow.

\subsection{Linear sources of peculiar velocities}\label{ssLSPVs}
Peculiar motions are believed to be an inevitable consequence of structure formation and of the resulting inhomogeneity and anisotropy of the universe. One can demonstrate this by simply recasting Eq.~(\ref{lrels2b}) as the linear evolution formula of the peculiar velocity field, namely as
\begin{equation}
\tilde{v}_a^{\prime}+ H\tilde{v}_a= \tilde{A}_a- A_a\,.  \label{tv'1}
\end{equation}
Accordingly, linear peculiar velocities are induced by differences between the two 4-acceleration vectors. Indeed, when $\tilde{A}_a- A_a=0$, the above reduces to
\begin{equation}
\tilde{v}_a^{\prime}= -H\tilde{v}_a\,,  \label{tv'2}
\end{equation}
guaranteing the $\tilde{v}_a^{\prime}=0$ when $\tilde{v}_a=0$. Put another way, in the absence of 4-acceleration sources, peculiar velocities remain zero if they were zero initially. Alternatively, assuming that $\tilde{v}_a\neq0$ to begin with, the above ensures that $\tilde{v}_a\propto a^{-1}$ on all scales. In other words, any peculiar velocity fields that might existed initially, will quickly decay away with the expansion. In such a case, the two relatively moving frames will start getting closer to each other until they eventually coincide. Overall, linear peculiar-velocity perturbations are not only generated, but they are also sustained by the presence of non-gravitational (in the relativistic sense) forces.

When the cosmic medium is a pressureless perfect fluid, we are free to set the 4-acceleration to zero in one coordinate system. Then, following (\ref{tv'1}), the 4-acceleration is nonzero in any other relatively moving frame. Technically speaking, when peculiar motions are present, one cannot simultaneously treat the worldlines of the relatively moving observers as timelike geodesics, even in the absence of pressure. For example, setting $A_a=0$ in the reference system, leads to
\begin{equation}
\tilde{v}_a^{\prime}+ H\tilde{v}_a= \tilde{A}_a\neq 0\,,  \label{tv'3}
\end{equation}
in the tilted frame of the real observers. In direct analogy, setting $\tilde{A}_a=0$ implies that
\begin{equation}
\dot{v}_a+ Hv_a= A_a\neq 0\,,  \label{vdot1}
\end{equation}
with $v_a=-\tilde{v}_a$ and the overdots indicating time derivatives in the reference frame. In either case, the source of the peculiar-velocity field is the 4-acceleration and the non-gravitational forces that it represents. Next, we will show that these non-gravitational forces are induced by structure formation and by the increasing inhomogeneity and anisotropy of the universe.

\subsection{The role of structure formation}\label{ssRSF}
One is free to chose either of Eqs.~(\ref{tv'3}) or (\ref{vdot1}) to describe the linear evolution of peculiar velocities in tilted almost-FRW universes. The results are the same, as expected, since physics is not frame dependent. Here, we will simply confirm this by taking both approaches.

Nonzero 4-acceleration reflects the presence of non-gravitational forces, which in our case are triggered by the ongoing process of structure formation. In order to make the connection, one needs to keep in mind that, when peculiar motions are present, there are nonzero energy fluxes due to relative-motion effects (at the linear level -- see Eq.~(\ref{lrels1}c) in \S~\ref{ssARM} previously). This means that a cosmic medium that appears perfect to the (idealised) Hubble-flow observers (i.e.~with $q_a=0$), will seem imperfect to their real counterparts in the tilted frame, since $\tilde{q}_a=-\bar{\rho}\tilde{v}_a$ (see (\ref{lrels1}c) and recall that we have set $p=0$). The reverse is also true, namely setting $\tilde{q}_a=0$ in Eq.~(\ref{lrels1}c) leads to $q_a=-\bar{\rho}v_a$ (with $v_a=-\tilde{v}_a$ to first order). Overall, in order to properly account for the effects of peculiar velocities, one needs to treat the cosmic fluid as imperfect, with an effective energy flux triggered by the relative motions. Following (\ref{tv'3}) and (\ref{vdot1}), the imperfect-fluid treatment is necessary irrespective of our frame choice.

Turning to relativistic linear cosmological perturbation theory, we linearise the nonlinear relations (1.3.18) and (2.3.1) of~\cite{TCM} -- or equivalently Eqs.~(5.12) and (10.101) of~\cite{EMM} -- around the tilted frame. The former of the above relations leads to $\tilde{A}_a=-(\tilde{q}_a^{\prime}+ 4H\tilde{q}_a)/\bar{\rho}$, which substituted into the latter provides the following linear expression for the 4-acceleration vector
\begin{equation}
\tilde{A}_a= {1\over3H}\,\tilde{\rm D}_a\tilde{\vartheta}- {1\over3aH}\left(\tilde{\Delta}_a^{\prime} +\tilde{\mathcal{Z}}_a\right)\,,  \label{tA}
\end{equation}
where $\tilde{\vartheta}=\tilde{\rm D}^a\tilde{v}_a$ (see also~\cite{TT} for further discussion and details). Alternatively, linearising expressions (1.3.18) and (2.3.1) of~\cite{TCM} in the idealised reference frame of the Hubble expansion, leads to $A_a=-(\dot{q}_a+ 4Hq_a)/\bar{\rho}$ and
\begin{equation}
A_a= {1\over3H}\,{\rm D}_a\vartheta- {1\over3aH}\left(\dot{\Delta}_a +\mathcal{Z}_a\right)\,,  \label{A}
\end{equation}
respectively (with $\vartheta={\rm D}^av_a=-\tilde{\vartheta}$). Note that $\tilde{\rm D}_a=\tilde{h}_a{}^b\nabla_a$ and ${\rm D}_a =h_a{}^b\nabla_b$ are the spatial covariant derivative operators in the tilted and the Hubble/CMB frames respectively (with $\tilde{h}_{ab}=g_{ab}+\tilde{u}_a\tilde{u}_b$ and $h_{ab}=g_{ab}+u_au_b$ being the associated projection tensors). Then, the 3-gradients $\tilde{\Delta}_a=(a/\bar{\rho})\tilde{\rm D}_a\rho$ and $\tilde{Z}_a=a\tilde{\rm D}_a\tilde{\Theta}$ describe linear spatial inhomogeneities in the matter density and in the universal expansion respectively~\cite{TCM,EMM}. In analogy, $\Delta_a=(a/\bar{\rho}){\rm D}_a\rho$ and $Z_a=a{\rm D}_a\Theta$ monitor the same perturbations in the Hubble/CMB frame. Note that all the perturbed variables vanish identically in the FRW background, which makes our linear study gauge invariant~\cite{SW}.

Combined, the sets (\ref{tv'3}), (\ref{vdot1}) and (\ref{tA}), (\ref{A}) show that peculiar motions are the inevitable result of the ongoing process of structure formation. In particular, substituting (\ref{tA}) into the right-hand side of (\ref{tv'3}), the latter takes the form
\begin{equation}
\tilde{v}_a^{\prime}+ H\tilde{v}_a= {1\over3H}\,\tilde{\rm D}_a\tilde{\vartheta}- {1\over3aH}\left(\tilde{\Delta}_a^{\prime} +\tilde{\mathcal{Z}}_a\right)\,,  \label{tv'4}
\end{equation}
guaranteeing that spatial gradients in the density distribution of the matter and in the expansion act as linear sources of peculiar velocities. The same conclusion also follows from expressions (\ref{vdot1}) and (\ref{tA}), which combine to give\footnote{Expression (\ref{tv'4}) also follows after linearising  Eqs.~(2.3.1) and/or (10.101), of~\cite{TCM} and~\cite{EMM} respectively, around the tilted frame and then substituting $\tilde{q}_a=-\bar{\rho}\tilde{v}_a$. Similarly, linearising the same formulae around the Hubble/CMB frame and then substituting $q_a=-\bar{\rho}v_a$ leads to Eq.~(\ref{vdot2}).}
\begin{equation}
\dot{v}_a+ Hv_a= {1\over3H}\,{\rm D}_a\vartheta- {1\over3aH}\left(\dot{\Delta}_a +\mathcal{Z}_a\right)\,.  \label{vdot2}
\end{equation}

The last pair of relations guarantees that linear peculiar velocities are generated by non-gravitational forces and, more specifically, by spatial gradients in the density distribution of the matter and by those in the universal expansion. Having established that, we will next proceed to analyse the linear evolution and the growth of cosmological peculiar-velocity fields.

\section{Peculiar velocities in pressureless matter}\label{sPVPM}
Density perturbations and peculiar velocities in the baryons start growing after recombination, once they have decoupled from the background photons. The CDM species, however, can develop their own density inhomogeneities and drift motions before matter-radiation equality.

\subsection{Peculiar velocities after recombination}\label{ssPVBAR}
Differentiating (\ref{tv'4}) in time, relative to the $\tilde{u}_a$-frame, employing the linear commutation law $(\tilde{\rm D}_a\tilde{\vartheta})^{\prime}=\tilde{\rm D}_a\tilde{\vartheta}^{\prime}+H\tilde{\rm D}_a\tilde{\vartheta}$ and then substituting Eq.~(\ref{tv'4}) back into the resulting expression, gives
\begin{equation}
\tilde{v}_a^{\prime\prime}+ \left(1-{1\over2}\,\Omega\right)H\tilde{v}_a^{\prime}- H^2(1+\Omega)\tilde{v}_a= {1\over3H}\,\tilde{\rm D}_a\tilde{\vartheta}^{\prime}- {1\over3aH} \left(\tilde{\Delta}_a^{\prime\prime} +\tilde{\mathcal{Z}}_a^{\prime}\right)\,,  \label{tv''1}
\end{equation}
with $\Omega=\kappa\bar{\rho}/3H^2$ being the background density parameter. Alternatively, one may substitute the time derivative of (\ref{tA}) into that of Eq.~(\ref{tv'2}), while using the aforementioned linear commutation law together with expression (\ref{tv'2}) again (see~\cite{TT} for the details). Relative to the Hubble frame, one  starts from Eq.~(\ref{vdot2}) and then proceeds in an exactly analogous way to arrive at
\begin{equation}
\ddot{v}_a+ \left(1-{1\over2}\,\Omega\right)H\dot{v}_a- H^2(1+\Omega)v_a= {1\over3H}\,{\rm D}_a\dot{\vartheta}- {1\over3aH} \left(\ddot{\Delta}_a+\dot{\mathcal{Z}}_a\right)\,,  \label{vddot1}
\end{equation}
to first approximation. Not surprisingly, Eqs.~(\ref{tv''1}) and (\ref{vddot1}) are formally identical, which ensures that they lead to the same results. Put another way, the linear evolution of the peculiar velocity field does not depend on the choice of the coordinate system. The idealised observers, namely those following the Hubble frame, reach the same conclusions with their real counterparts in the tilted frame, once the relative motion of the two coordinate systems has been accounted for.

Equations (\ref{tv''1}) and (\ref{vddot1}) monitor the linear evolution of peculiar velocities in pressureless matter in a tilted almost-FRW universe. The former holds in the tilted frame of the real observes and the latter in that of their Hubble-flow counterparts respectively. One may therefore apply the above to baryonic ``dust'' after recombination and to low-energy CDM species before, as well as after, decoupling.

Before proceeding to the solution of (\ref{tv''1}) and/or (\ref{vddot1}), certain technical comments are in order. Both expressions are inhomogeneous differential equations, of which only the homogeneous component is analytically solvable.\footnote{It should be noted that the inhomogeneity/homogeneity of (\ref{tv''1}) and (\ref{vddot1}) refers only to the nature of these differential equations and not to the inhomogeneity/homogeneity of the host spacetime.} We will therefore only solve the homogeneous parts of (\ref{tv''1}) and (\ref{vddot1}). Then, strictly speaking, our results apply to relatively large scales, where the spatial gradients seen on the right-hand side of these two expressions are expected to be subdominant. Having said that, the full solution of an inhomogeneous differential equation is formed by the full solution of its homogenous component and by a partial solution of the inhomogeneous equation. In practice, this means that solving the full differential equation will not make any real physical difference, unless the partial solution grows faster than the faster growing mode of the homogeneous solution. Put another way, if solving the homogeneous components of (\ref{tv''1}), (\ref{vddot1}) leads to a growing mode for the peculiar velocity field, the solution of the full equations can only increase the growth rate. Note, however, that in the (fairly unlikely) case the partial solution grows as fast as the fastest growing homogeneous mode, the outcome of Eqs.~(\ref{tv''1}), (\ref{vddot1}) may not be as straightforward.

The homogeneous component of (\ref{tv''1}) was solved in~\cite{TT}, assuming an Einstein-de Sitter background universe with $\Omega=1$ and $H=2/3t$. Indeed, written on such a background, the homogeneous parts of Eqs.~(\ref{tv''1}) and (\ref{vddot1}) reduce to
\begin{equation}
\tilde{v}_a^{\prime\prime}+ {1\over3t}\,\tilde{v}_a^{\prime}- {8\over9t^2}\,\tilde{v}_a= 0 \hspace{10mm} {\rm and} \hspace{10mm} \ddot{v}_a+ {1\over3t}\,\dot{v}_a- {8\over9t^2}\,v_a= 0\,,  \label{bvs}
\end{equation}
respectively. Both of the above solve analytically, giving $\tilde{v},\,v\propto t^{4/3}\propto a^2$, which is considerably stronger than the $v\propto t^{1/3}\propto a^{1/2}$ of the Newtonian (or quasi-Newtonian) analysis. The above solution applies to both baryons and CDM, after recombination and equilibrium respectively. Note that, due to the nature of the differential equations involved, this growth rate might increase further if the inhomogeneous components of (\ref{tv''1}), (\ref{vddot1}) are accounted for (see earlier comments).

\subsection{Peculiar velocities in CDM prior to
recombination}\label{ssPVCDMPR}
Prior to recombination density perturbations in the baryonic sector cannot grow. During the radiation era, for example, the Jeans length of the (tightly coupled) photon-baryon system matches that of the horizon. Also, between equilibrium and decoupling photon diffusion, from overdense to underdense regions, leads to Silk damping and effectively erases any inhomogeneities that might existed in the baryon density (see~\cite{S} as well as~\cite{KT,CL}). Given that peculiar velocities are triggered by inhomogeneities in the matter distribution, we do not expect substantial velocity perturbations to develop in the baryonic component. Typical CDM species, on the other hand, do not interact neither with the photons nor with the baryons and they become non-relativistic during the radiation epoch. More specifically, the heavier the dark matter candidate, the earlier it becomes non-relativistic. In addition, the free-streaming (or Landau damping) scale of the CDM particles is typically too small to have cosmological relevance. Nevertheless, perturbations in the CDM density cannot grow during most of the radiation era due to the Mezsaros stagnation effect (see~\cite{Me}, as well as~\cite{TCM,EMM}). It is only near the time of matter-radiation equality that linear inhomogeneities in the dark-matter sector can start to grow and, in the process, trigger peculiar-velocity perturbations in their own distribution. Given that low-energy CDM behaves dynamically like pressure-free dust, Eqs.~(\ref{tv''1}), (\ref{vddot1}) can also be used to monitor the linear evolution of peculiar velocities in the dark matter during the radiation era, when $H=1/2t$. In so doing, we assume that photon diffusion keeps both the radiation field and the (tightly coupled) baryons homogeneously distributed throughout that epoch, which is a reasonable approximation to make at least on subhorizon scales.

Isolating the homogeneous component of differential equation (\ref{tv''1}) and assuming spatial flatness (i.e.~setting $\Omega=1$ and $H=1/2t$), the latter reads
\begin{equation}
\tilde{v}_a^{\prime\prime}+ {1\over4t}\,\tilde{v}_a^{\prime}- {1\over2t^2}\,\tilde{v}_a= 0\,,  \label{CDMv''1}
\end{equation}
which solves to give
\begin{equation}
\tilde{v}= \mathcal{C}_1t^{(3+\sqrt{41})/8}+ \mathcal{C}_2t^{(3-\sqrt{41})/8}= \mathcal{C}_3a^{(3+\sqrt{41})/4}+ \mathcal{C}_4a^{(3-\sqrt{41})/4} \,.  \label{CDMv}
\end{equation}
Therefore, for a brief period prior to equipartition, linear peculiar-velocity perturbations in the CDM component also grow, though at a time-rate slower than after matter-radiation equality (compare the above to solution (\ref{bvs}) in the previous section).

After recombination, the peculiar-velocity fields that have already developed in the dark matter sector could also enhance those in the baryonic component, in the same way that ``gravitational wells'' in the CDM distribution amplify density perturbations in the baryons.\footnote{Studies involving peculiar velocities in the CDM component, though without focusing on their linear evolution and growth, have also appeared in the recent literature. For instance, drift motions in the CDM sector, within Szekeres-type regions matched on a $\Lambda$CDM background, have been investigated in an attempt to alleviate the Hubble-tension problem~\cite{NS}. Also, studies of dark-matter decay in cosmic voids, involving tilted cosmological models, were recently pursued in~\cite{LB}.}

\section{Peculiar velocities during inflation}\label{sPVDI}
Large-scale peculiar motions are generally believed to be a recent addition to the kinematics of our universe and a byproduct of structure formation. Having said that, it is conceivable and it has been suggested that peculiar velocities might instead have primordial origin~\cite{T}.

\subsection{The dynamics of de Sitter inflation}\label{ssDdSI}
Typical inflationary scenarios assume that the very early universe is dominated by scalar fields, like those predicted by the theories of high-energy physics. When these scalar fields are minimally coupled to gravity, they behave as effective perfect fluids with energy-momentum tensor
\begin{equation}
T_{ab}= \bar{\rho}u_au_b+ \bar{p}h_{ab}\,,  \label{Tab}
\end{equation}
relative to the timelike 4-velocity $u_a$~\cite{Ma}. The effective energy density and pressure of the inflaton field ($\phi$) are respectively given by
\begin{equation}
\bar{\rho}= {1\over2}\,\dot{\phi}^2+ V \hspace{15mm} {\rm and} \hspace{15mm} \bar{p}= {1\over2}\,\dot{\phi}^2- V\,,  \label{phi}
\end{equation}
where $V=V(\phi)$ is the associated potential and the overdots indicate time derivatives relative to the $u_a$-field. Another key relation is the Klein-Gordon equation, which monitors the time evolution of $\phi$ and is given by
\begin{equation}
\ddot{\phi}+ 3H\dot{\phi}+ \hat{V}= 0\,,  \label{KG}
\end{equation}
with $\hat{V}={\rm d}V/{\rm d}\phi$~\cite{TCM,EMM}.

In order to achieve exponential expansion of the de Sitter type, one needs to impose the slow-rolling conditions, namely demand that $\dot{\phi}=0$ and $3H\dot{\phi}+\hat{V}=0$ simultaneously. The former condition ensures that the inflaton field rolls slowly down its potential, which in turn guarantees that $\bar{p}=-\bar{\rho}=-V$ and therefore exponential expansion. The latter condition ensures that $\ddot{\phi}=0$ and thus guarantees that the slow-rolling phase lasts long enough to bring the primordial universe (arbitrarily) close to a spatially flat radiation-dominated Friedmann model.

\subsection{Tilted inflationary models}\label{ssTIMs}
Consider a perturbed inflationary model and allow for a pair of relatively moving frames as described in \S~\ref{ssARM}. Assuming that the peculiar velocity between the two coordinate systems is non-relativistic, the linear relations (\ref{lrels1}) still apply. Nevertheless, there is a key difference between tilted inflationary models and those containing conventional matter, because $\bar{p}=-\bar{\rho}$ during the de Sitter regime. Applied to expression (\ref{lrels1}c), the equation of state of a slow-rolling scalar field ensures that $\tilde{q}_a=q_a$ throughout the exponential expansion phase. This result, together with relation (\ref{lrels1}d) implies that, if the (effective) cosmic fluid appears perfect in one set of observers, it will look the same in any other relatively moving coordinate system as well.

The linear kinematic relations (\ref{lrels2a}) and (\ref{lrels2b}) also apply to the de Sitter phase. In fact the latter relation guarantees that non-gravitational forces are the sources of peculiar velocities during inflation as well. Following~\cite{TCM,EMM}, the 4-acceleration vectors representing the aforementioned forces are $A_a=-({\rm D}_a\dot{\phi)}/\dot{\phi}$ in the reference Hubble/CMB frame and $\tilde{A}_a=-(\tilde{\rm D}_a\phi^{\prime})/ \phi^{\prime}$ in its tilted counterpart.\footnote{Both $\phi$ and $V=V(\phi)$ are spacetime quantities and therefore do not change after a Lorentz boost between frames. In other words, $\tilde{\phi}=\phi$ and $\tilde{V}=V$ at all perturbative levels.} Therefore, as long as $\tilde{A}_a-A_a\neq0$, peculiar-velocity perturbations can in principle survive a phase of exponential inflation. Let us take a closer look at this possibility.

\subsection{Peculiar velocities during inflation}\label{ssPVDI}
During the de Sitter phase, the linear evolution of primordial peculiar-velocity perturbations is monitored by (\ref{tv'1}), which in our inflationary environment acquires the form
\begin{equation}
\tilde{v}_a^{\prime}+ H\tilde{v}_a= {1\over\phi^{\prime}}\,\tilde{\rm D}_a\phi^{\prime}- {1\over\dot{\phi}}\,{\rm D}_a\dot{\phi}\,.  \label{dStv'1}
\end{equation}
Although $\phi$ and $V=V(\phi)$ remain invariant under our frame change (see footnote~4), this is not a priori guaranteed for their time derivatives and for their spatial gradients. More specifically, $\phi^{\prime}\neq\dot{\phi}$ and $\tilde{\rm D}_a\phi^{\prime}\neq{\rm D}_a\dot{\phi}$ in general. Having said that, confining to the linear regime, we find
\begin{equation}
\phi^{\prime}= \tilde{u}^a\nabla_a\phi= \dot{\phi}+ \tilde{v}^a{\rm D}_a\phi= \dot{\phi}\,,  \label{phis1}
\end{equation}
given that $\tilde{v}^a{\rm D}_a\phi$ is second order perturbatively. Similarly, using the above result and keeping in mind that $\tilde{h}_{ab}=h_{ab}+2u_{(a}\tilde{v}_{b)}$ to first approximation -- see~\cite{M}, we arrive at the linear relation
\begin{equation}
\tilde{\rm D}_a\phi^{\prime}= {\rm D}_a\dot{\phi}+ \ddot{\phi}\tilde{v}_a= {\rm D}_a\dot{\phi}\,,  \label{phis2}
\end{equation}
since $\ddot{\phi}=0$ throughout the slow-rolling phase (see Eq.~(\ref{KG}) and discussion immediately after).\footnote{Strictly speaking, the slow-roll conditions are $\dot{\phi}\simeq0\Leftrightarrow p-\rho\simeq0$ and $\ddot{\phi}\simeq0$. These ensure that the peculiar velocity terms (i.e.~$(p-\rho)\tilde{v}_a$ and $\ddot{\phi}\tilde{v}_a$ respectively) seen in Eqs.~(\ref{lrels1}c) and (\ref{phis2}) are (for all practical purposes) of the second perturbative order, which in turn guarantees the consistency and the generality of our linear analysis.}

Combining relations (\ref{dStv'1}), (\ref{phis1}) and (\ref{phis2}) the former of the three ``loses'' its right-hand side and reduces to the source-free linear propagation formula
\begin{equation}
\tilde{v}_a^{\prime}+ H\tilde{v}_a= 0\,.  \label{dStv'2}
\end{equation}
Recalling that $H=H_0=$~constant during de Sitter inflation, the above immediately gives
\begin{equation}
\tilde{v}_a\propto a^{-1}\propto {\rm e}^{-H_0t}\,,  \label{dStv}
\end{equation}
with the zero suffix indicating the start of the exponential expansion phase. Therefore, during de Sitter inflation linear peculiar velocities decay exponentially. Following solution (\ref{dStv}), the peculiar velocity at the end of the de Sitter era is
\begin{equation}
\tilde{v}_f= \tilde{v}_0{\rm e}^{-\mathcal{N}}\,,  \label{dSvf}
\end{equation}
where $\mathcal{N}$ is the number of e-folds. Thus, assuming that the exponential expansion phase lasts for 60 e-folds, which is the minimum required period for inflation to work, the peculiar-velocity field drops by approximately 26 orders of magnitude. For 100 e-fold inflation, the decrease is roughly 43 orders of magnitude and so on.

In practice, any 4-velocity ``tilt'', which might had been present at the onset of inflation, would have quickly decayed away by the end of the de Sitter era, leaving the post-inflationary universe free of peculiar-velocity perturbations. This result reflects the fact that $\tilde{A}_a=A_a$ throughout the de Sitter phase, which in turn ensures that the peculiar-velocity field remains source-free at the linear level. The absence of sources from Eq.~(\ref{tv'1}) is a direct consequence of the inflationary equation of state (i.e.~$p=-\rho$), which ensures that (to linear order) the energy-flux vector vanishes in all relatively moving frames, when it happens to be zero in one of them (i.e.~$\tilde{q}_a=q_a=0$ -- see relation (\ref{lrels1}c) in \S~\ref{ssARM}).

\section{Discussion}\label{sD}
The origin, the evolution, the role and the implications of the bulk peculiar flows reported by the related surveys are still largely unknown. Large-scale peculiar velocities are believed to be a recent addition to the kinematics of our universe, though the possibility of a primordial origin for them has been raised as well. Relative motions have been typically associated with the presence of apparent dipole-like anisotropies, like the one seen in the CMB spectrum (e.g.~see~\cite{Ketal}). It is therefore conceivable that some part (at least) of the dipolar axes reported in the recent literature (e.g.~see~\cite{CMRS}-\cite{Cetal}), may be artifacts of our peculiar motion relative to the Hubble expansion (e.g.~see~\cite{T1}). All these issues remain largely open and a subject of debate. This may have to do with the data, which remain patchy and rather difficult to extract, but it may also reflect the fact that the theoretical studies are still few and sparse. Our work approached the issue from the relativistic viewpoint, by applying linear cosmological perturbation theory to a tilted almost-FRW universe. The latter allows for two families of relatively moving observers and therefore provides the appropriate theoretical environment to study peculiar motions in cosmology. It goes without saying that the results are independent of the adopted reference frame, namely of whether we take the view point of the real (tilted) observers, or of their idealised (Hubble-flow) counterparts.

Our first step was to confirm that, at the linear level, peculiar velocities are triggered by non-gravitational forces, which in turn are induced by the ongoing structure formation process. Technically speaking, within the relativistic framework, linear peculiar velocities are generated and sustained by a nonzero net 4-acceleration vector (see Eq.~(\ref{tv'1}) and also (\ref{tv'2}), (\ref{vdot1}) in \S~\ref{ssLSPVs}). In Newtonian theory, the latter is expressed in terms of the gravitational potential, leading to a growth rate of $v\propto t^{1/3}$ after recombination. There is no gravitational potential in relativity. There, the 4-acceleration is directly related to linear inhomogeneities in the matter density and in the universal expansion, even in the absence of pressure (see expressions (\ref{tA}) and (\ref{A}) in \S~\ref{ssRSF}). This is the key difference separating the Newtonian from the relativistic analysis of linear peculiar motions and the reason is the flux input of the peculiar flow to the energy-momentum tensor and therefore to the local gravitational field. There is no such flux contribution in Newton's gravity by default. As a result, the relativistic evolution formulae of the peculiar velocity field (see Eqs.~(\ref{tv'4}), (\ref{vdot2}) in \S~\ref{ssRSF}) cannot be naturally reproduced by the Newtonian studies, which explains why the two treatments reach different results and conclusions. More specifically, the relativistic study of linear
peculiar velocities in the pressureless component of the matter (baryonic or dark), lead to a (considerably stronger) growth rate of $v\propto t^{4/3}$ after recombination~\cite{TT}. Here, we confirmed these results and also extended the analysis to earlier epochs in the lifetime of the universe.

Prior to recombination, baryonic aggregations are not allowed to grow gravitationally, which means that the baryons remain homogeneously distributed (for all practical purposes) at least on subhorizon scales. This in turn should prevent peculiar velocities from developing in the baryonic sector. Although the CDM species become non-relativistic well before equipartition, density perturbations and therefore peculiar velocities in the dark-matter sector can start growing only in the late radiation era. We found that such velocity perturbations grow as $v\propto t^{(3+\sqrt{41})/8}$ at the linear level, a growth-rate slightly slower than that of the subsequent dust era. Then, if peculiar velocities already exist in the CDM component at the time of recombination, they can in principle amplify those in the baryons, in the same way the potential wells in the dark matter distribution enhance the growth of baryonic density perturbations after decoupling.

In all the scenarios of peculiar-velocity evolution discussed above, the decisive common factor is the peculiar flux contribution to the local gravitational field. The flux effect, which emerges naturally in tilted cosmological models, is purely relativistic, it is triggered by the peculiar motion of the matter and occurs at the linear level. As a result, the relativistic analysis enhances the linear growth-rate of peculiar-velocity perturbations, compared to the Newtonian treatment. The pivotal role played by the aforementioned flux contribution becomes evident when considering peculiar velocities during de Sitter inflation. In typical inflationary scenarios, the unconventional ($\bar{p}=-\bar{\rho}$) equation of state of the slow-rolling inflaton, guarantees that the peculiar energy-flux vanishes in all relatively moving frames (at the linear level -- see relation (\ref{lrels1}c) in \S~\ref{ssARM}). In the absence of any flux contribution to the local gravitational field, linear peculiar velocities were found to decay exponentially throughout the de Sitter phase. Therefore, even if there was a primordial peculiar-velocity ``tilt'' at the onset of inflation, it should vanish by the end of it. Having said that, it is conceivable that there could be inflationary scenarios that allow peculiar velocities to survive.

Overall, the relativistic analysis seems to favour faster peculiar velocities in both the baryonic and the CDM component and therefore it appears to support reports of bulk flows faster than it has been typically anticipated. During standard de Sitter inflation, on the other hand, any peculiar-velocity perturbations that might have already existed are exponentially suppressed. Therefore, according to our study, it seems rather unlikely that the peculiar-velocity fields observed in the universe today could have primordial origin.\\

\textbf{Acknowledgements:} This work was supported by the Hellenic Foundation for Research and Innovation (H.F.R.I.), under the ``First Call for H.F.R.I. Research Projects to support Faculty members and Researchers and the procurement of high-cost research equipment Grant'' (Project Number: 789).\\

\end{document}